\documentclass[pra,  twocolumn, superscriptaddress]{revtex4-2}
\def\be{\begin{equation}}
\def\ee{\end{equation}}
\def\bea{\begin{eqnarray}}
\def\eea{\end{eqnarray}}
\def\bi{\begin{itemize}}
\def\ei{\end{itemize}}
\usepackage{braket}

\usepackage{xcolor}
\usepackage{graphicx}
\usepackage{amsmath,amstext,amssymb,bm,color,times}
\usepackage[english]{babel}
\usepackage[T1]{fontenc}
\usepackage[utf8]{inputenc}
\usepackage[colorlinks=true,citecolor=blue,linkcolor=magenta]{hyperref}
\usepackage{lipsum}

\begin{document}


\title{
Inhomogeneous adiabatic preparation of a quantum critical ground state in two dimensions
}

\author{Ihor Sokolov}
\affiliation{Jagiellonian University, Institute of Theoretical Physics, {\L}ojasiewicza 11, PL-30348 Krak\'ow, Poland}
\author{Francis A. Bayocboc Jr.}
\affiliation{Jagiellonian University, Institute of Theoretical Physics, {\L}ojasiewicza 11, PL-30348 Krak\'ow, Poland}
\author{Marek M. Rams}
\affiliation{Jagiellonian University, Institute of Theoretical Physics, {\L}ojasiewicza 11, PL-30348 Krak\'ow, Poland}
\author{Jacek Dziarmaga}
\affiliation{Jagiellonian University, Institute of Theoretical Physics, {\L}ojasiewicza 11, PL-30348 Krak\'ow, Poland}

\date{\today}

\begin{abstract}
Adiabatic preparation of a critical ground state is hampered by the closing of its energy gap as the system size increases.
However, this gap is directly relevant only for a uniform ramp, where a control parameter in the Hamiltonian is tuned uniformly in space towards the quantum critical point. 
Here, we consider inhomogeneous ramps in two dimensions: 
initially, the parameter is made critical at the center of a lattice, from where the critical region expands at a fixed velocity.
In the 1D and 2D quantum Ising models, which have a well-defined speed of sound at the critical point, the ramp becomes adiabatic with a subsonic velocity. This subsonic ramp can prepare the critical state faster than a uniform one.
Moreover, in both a model of $p$-wave paired 2D fermions and the Kitaev model, the critical dispersion is anisotropic---linear with a nonzero velocity in one direction and quadratic in the other---but the gap is still inversely proportional to the linear size of the critical region, with a coefficient proportional to the nonzero velocity. This suffices to make the inhomogeneous ramp adiabatic below a finite crossover velocity and superior to the homogeneous one.
\end{abstract}
\maketitle


\section{Introduction}
\label{sec:intro}

The idea of an adiabatic quantum state preparation (AP) is to start from an easy-to-prepare ground state of an initial Hamiltonian $H_i$ and then drive the Hamiltonian adiabatically to a final Hamiltonian $H_f$, whose ground state is the desired state~\cite{RevModPhys_Lidar_AQC}. This target state may contain a solution to a hard computational problem that only needs to be measured in a suitable basis, making AP identical to adiabatic quantum computation~\cite{RevModPhys_Lidar_AQC,King_Dwave1d_2022}. Alternatively, the ground state of $H_f$ may be interesting in its own right with all its correlations, aligning AP more with quantum simulation. The Hamiltonian is typically ramped as $H(\lambda)=(1-\lambda) H_i+\lambda H_f$, with the parameter $\lambda$ smoothly increasing from $0$ to $1$ within a ramp time $\tau$. Adiabaticity can be enhanced by optimizing the ramp profile $\lambda(t)$, slowing its increase rate where the gap is smaller, or by adding counter-adiabatic terms to $H(\lambda)$ in order to make a shortcut to adiabaticity~\cite{del_Campo_2019,keever2023adiabatic}.

AP becomes hard when the gap closes at a quantum critical point. For instance, driving $H(\lambda)$ across a critical point is described by the quantum version~\cite{QKZ1,QKZ2,QKZ3,d2005,d2010-a, d2010-b, KZnum-g,QKZteor-b,QKZteor-c,QKZteor-d,QKZteor-e,QKZteor-f,QKZteor-g,QKZteor-h,Barankov_nonlinKZ,PRBnonlinKZ,QKZteor-i,QKZteor-j,QKZteor-k,QKZteor-l,QKZteor-m,QKZteor-n,QKZteor-o,KZLR1,KZLR2,QKZteor-oo,delcampostatistics,KZLR3,QKZteor-q,QKZteor-r,QKZteor-s,QKZteor-t,sonic,QKZteor-u,QKZteor-v,QKZteor-w,QKZteor-x,roychowdhury2020dynamics,sonic, schmitt2021quantum,RadekNowak,dziarmaga_kinks_2022,transverse_oscillations,QKZexp-a, QKZexp-b, QKZexp-c, QKZexp-d, QKZexp-e, QKZexp-f, QKZexp-g,deMarco2,Lukin18,adolfodwave,2dkzdwave,King_Dwave1d_2022,King_Dwave_glass,king2024computational,miessen2024benchmarking} of the Kibble-Zurek mechanism (KZM)~\cite{K-a, *K-b, *K-c,Z-a,*Z-b,*Z-c,*Z-d}, which predicts the density of excitations (or excitation energy) to decay with a small power of the ramp time. At least for a symmetry-breaking transition, adding a tiny symmetry-breaking bias could open a gap when crossing the critical point and suppress excitations exponentially~\cite{QKZteor-r,KZexp-x}, but any such bypass of the criticality becomes useless when the target is the critical state itself.

\begin{figure}[t!]
\begin{centering}
\includegraphics[width=0.999\columnwidth]{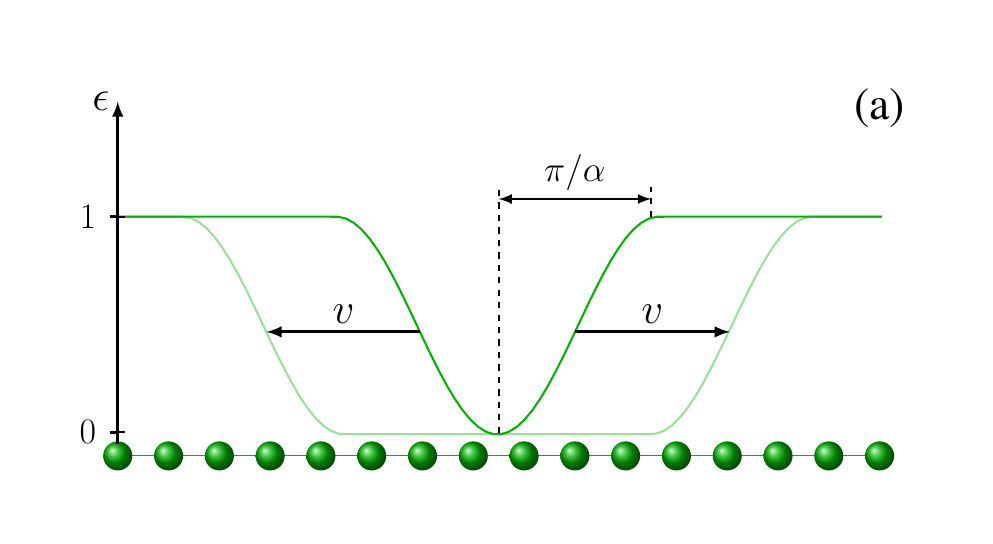}
\includegraphics[width=0.999\columnwidth]{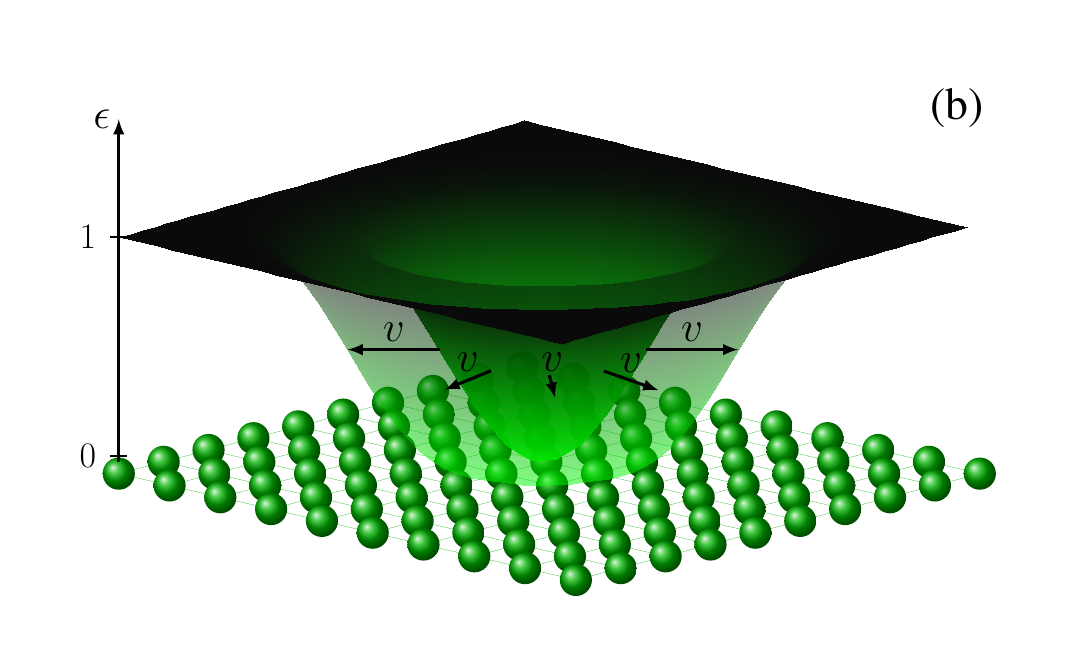}
\par\end{centering}
\caption{ 
{\bf Inhomogeneous ramp. }
Here, $\epsilon$ is a dimensionless parameter in a Hamiltonian that measures the distance to the quantum critical point.
(a) shows an inhomogeneous ramp in a 1D chain at two different times. The transition begins at the center of the chain, where the critical $\epsilon=0$ is reached first, and from where it expands with velocity $v$ towards the chain's ends. 
(b) shows an inhomogeneous ramp in a $10\times 10$ square lattice. The critical region with $\epsilon=0$ expands with velocity $v$ from the center towards the edges of the lattice. 
Both (a) and (b) demonstrate ramps with slope $\alpha=1$.
}
\label{fig:inhomo-artist's}
\end{figure}

When the critical $H_f$ is approached uniformly in space, the AP ramp is a ``half-KZM'' ramp, ending at the critical point instead of crossing it, and the KZM power laws still apply, as discussed at length in the following Sec.~\ref{sec:uniform}.  
A full KZM ramp can be made adiabatic by making it inhomogeneous in space~\cite{inhomo_quantum-a,inhomo_quantum-aa,inhomo_quantum-b,inhomo_quantum-c,inhomo_quantum-d,inhomo_quantum-e,inhomo_quantum-g,inhomo_quantum-h,inhomo_quantum-i,inhomo_quantum-j,inhomo_quantum-k}. Initially, the parameter $\lambda(t,{\bf r})$ can be driven across the critical point at the center of the lattice, then the central region with $H_f$ is expanded towards the lattice edges with a velocity $v$ slower than the speed of sound at the critical point, see Fig.~\ref{fig:inhomo-artist's}. 

An inhomogeneous ramp has also been considered as a means to prepare a critical ground state, either in a 1D chain~\cite{inhomo_quantum-f} or a quasi-1D stripe~\cite{bernier2024spatiotemporal}. These studies focused on slightly supersonic ramps that create excitations, which follow the ramp towards the ends of the chain, leaving behind the critical ground state in the bulk.
Since the excitations eventually bounce back from the ends, we prefer to focus on subsonic ramps in this work. Furthermore, instead of using excitation energy as a cost function---which under-represents low-frequency excitations that can distort the target state without significantly increasing its energy---we focus on the density of excited quasiparticles (for integrable systems) or, more generally, on the (in)fidelity between the critical ground state and the final state (for nonintegrable systems).

\section{A uniform ramp to a critical point}
\label{sec:uniform}

In this section, we consider ramping a system towards its quantum critical point by continuously varying a parameter in its Hamiltonian towards its critical value uniformly in space. This is a half quantum Kibble-Zurek ramp that, instead of crossing the quantum phase transition, terminates at the critical point. In line with the quantum KZM, instead of the bare Hamiltonian parameter, it is more convenient to use a dimensionless parameter $\epsilon$ that measures the distance to the critical point and is turned down to zero as
\be 
\epsilon(t) = \left| \frac{t}{\tau} \right|^r.
\label{eq:epsilon_r}
\ee 
Here, time $t\leq0$, $\tau$ is the ramp time, and $r>0$ is the ramp exponent. The power law~\eqref{eq:epsilon_r} is assumed only near the end of the ramp, close to the critical point where $\epsilon$ is small. Before the ramp begins at a large initial $\epsilon$, the system is initialized in an easy-to-prepare ground state of the initial Hamiltonian. Thanks to its large energy gap, the initial stage of the evolution is adiabatic. The adiabaticity must fail when the critical point is approached and the gap shrinks to zero. For a long enough ramp time $\tau$ the failure happens close enough to the critical point for the asymptote~\eqref{eq:epsilon_r} to be accurate and, furthermore, for the gap to scale as $\Delta\propto|\epsilon|^{z\nu}$, where $z$ and $\nu$ are the dynamical and correlation length exponents, respectively. According to the nonlinear KZM~\cite{QKZteor-h,PRBnonlinKZ,Barankov_nonlinKZ}, see also the review~\cite{d2010-a}, the failure happens when the rate of the transition, $|\dot\epsilon/\epsilon|\propto 1/|t|$, becomes comparable to the gap near $-\hat t$, where
\be 
\hat t \propto \tau^{r z\nu/(1+r z\nu)}.
\label{eq:hatt}
\ee 
In first approximation, after $-\hat t$, the evolution becomes impulse, i.e., the state freezes out because the Hamiltonian is varying too fast for the state to catch up with it. In this way the ramp ends at $t=0$ with the state approximately equal to the ground state at $\hat\epsilon=(\hat t/\tau)^r$ with a correlation length
\be 
\hat\xi \propto \hat\epsilon^{-\nu} \propto \tau^{r\nu/(1+r z\nu)}.
\label{eq:hatxi}
\ee 
As discussed at length in Ref.~\onlinecite{sonic}, the adiabatic-impulse approximation is not quantitatively accurate, as the correlation range can more than double between $-\hat t$ and $0$. However, it correctly predicts the power law~\eqref{eq:hatxi} satisfied by the KZ length $\hat\xi$, which is the unique scale of length characterizing the state in the framework of the KZ scaling hypothesis~\cite{KZscaling1,KZscaling2,Francuzetal}. In particular, the density of pointlike excitations scales as
\be 
\rho \propto \hat\xi^{-d} \propto \tau^{-d r\nu/(1+r z\nu)},
\label{eq:rho}
\ee 
where $d$ is the lattice dimensionality. The inverse,
\be 
\hat k \propto \hat\xi^{-1} \propto \tau^{-r\nu/(1+r z\nu)},
\label{eq:hatk}
\ee 
sets a scale for the maximal quasimomentum of excited quasiparticles~\cite{Francuzetal}. With the quasiparticle dispersion, $\omega\propto|k|^z$, the maximal group velocity of excited quasiparticles is
\be 
\hat v \propto \hat k^{z-1} \propto \tau^{-r(z-1)\nu/(1+r z\nu)}.
\label{eq:hatv}
\ee 
When $z=1$, then $\hat v$ is the speed of sound, but in general, the speed limit depends on the ramp time. For $z>1$, the excited quasiparticles have velocities ranging from $0$ for small $k$ to $\hat v$ near $\hat k$. 

Most importantly for the adiabatic preparation, the KZ length $\hat\xi$ discriminates between the KZM regime, where $\hat\xi\ll L$ and the above power laws hold, and the adiabatic regime, where $\hat\xi\gg L$. In the latter, the evolution remains adiabatic up to the critical point thanks to a finite gap due to the finite size of the system. The number of excitations decays exponentially with the ramp time and can be often described by a Landau-Zener process~\cite{d2005}. The condition $\hat\xi\gg L$ translates to ramp times
\be 
\tau \gg L^{z+\frac{1}{r\nu}}.
\label{eq:tau_adiab}
\ee 
The right hand side is the minimal time necessary to make the uniform ramp adiabatic.

The limit $r\to\infty$ might appear optimal for the adiabaticity, but it drives~\eqref{eq:epsilon_r} to zero, and the adiabatic-impulse crossover can happen within its range of applicability for extremely long ramp times only.
In order to make the point more precise without juggling with the limits, let us consider a ramp 
\be 
\epsilon(t) \propto \exp(\tau/t)
\label{eq:eps_exp}
\ee
that approaches $0$ faster than for any finite $r$. The transition rate $|\dot\epsilon/\epsilon|=\tau/t^2$ equals the gap $\Delta\propto\epsilon^{z\nu}$ when $\tau \hat t^{-2}=\exp(z\nu\tau/\hat t)$. A solution for $\hat t$ involves the Lambert function whose asymptote yields $\hat\epsilon\propto(\tau\ln^2\tau)^{-1/(z\nu)}$ and the KZ length
\be 
\hat\xi \propto \hat\epsilon^{-\nu} \propto (\tau\ln^2\tau)^{1/z}.
\ee 
Up to a logarithmic correction, it is consistent with the limit $r\to\infty$ of~\eqref{eq:hatxi}. However, the asymptote is approached very slowly and requires not only the usual $\tau\gg 1$ but even $\ln\tau\gg1$ or, equivalently, the ramp time spanning many orders of magnitude. On the one hand, for extremely long $\tau$, the adiabatic-impulse crossover happens at an exponentially small $\epsilon$ that is difficult to control experimentally. On the other hand, for a relatively short $\tau$, the ramp~\eqref{eq:eps_exp} can be caricatured by a linear ramp shrinking to $\epsilon=0$ near $t\approx -0.15\tau$, resulting in an effective $r_{\rm eff}=1$. Thus, for $\tau$ spanning one or two orders of magnitude, $\hat\xi$ appears to satisfy the power law~\eqref{eq:hatxi} with $r_{\rm eff}\in(1,\infty)$ depending on the range of $\tau$. This is similar to the KZM in the Kosterlitz-Thouless transition~\cite{KZM1DBHM,KZM1DBHM_Gardas}.

In order to avoid these traps we employ a simple symmetric ramp profile with a timelike parameter $u$:
\begin{equation}
\epsilon(u) =  \left\{ 
\begin{array}{ccc}
1, & \mathrm{for} & u \le -\pi/2,   \\
\frac12 - \frac12 \sin u , & \mathrm{for} & |u| < \pi/2, \\
0 & \mathrm{for} & u \ge \pi/2. 
\end{array} 
\right.
\label{eq:eps_cr}
\end{equation}
This profile has a continuous time derivative at $u=\mp\pi/2$ to suppress initial and final excitations. For the uniform transition, we set
\be 
u=t/\tau.
\label{eq:u_uni}
\ee
The ramp approaches its end at $u=\pi/2$ in a parabolic way with $r=2$, which seems to be a reasonable compromise between a linear ramp and the large $r$ limit. For later comparison with inhomogeneous transitions, we note that
\be 
\tau_{total} = \pi\tau 
\label{eq:tau_tot_uni}
\ee 
is the total time to complete the uniform ramp.

\section{An inhomogeneous ramp to a critical point}
\label{sec:inhomo}

We consider inhomogeneous ramps that employ the envelope function~\eqref{eq:eps_cr} with a local timelike parameter
\be 
u = \alpha \left[ v t - d(s, s_c) \right].
\label{eq:u_inhomo}
\ee
Here, $v$ is the spatial velocity of the ramp that we assume to be time-independent, $\alpha$ is its slope, and $d(s,s_c)$ is the distance between site $s$ and the lattice center $s_c$. For a NN interaction, $s$ is the center of a NN bond. The transition begins at $s_c$, where $\epsilon=0$ is reached first, and from where the region with the critical Hamiltonian expands with velocity $v$ towards the edges of the lattice (see Fig.~\ref{fig:inhomo-artist's}). For an $L\times L$ square lattice, it takes a total time
\be 
\tau_{total} = \frac{\pi}{\alpha v} + \frac{1}{v} \frac{L-1}{\sqrt2}
\label{eq:tau_tot_inhomo}
\ee 
to fully execute the ramp everywhere between its center and four corners. The offset $\frac{\pi}{\alpha v}$ is due to the finite width of the ramp. This is the time needed for $\epsilon$ to reach $0$ at $s_c$ plus the time from when $\epsilon$ starts decreasing until it becomes $0$ at the corners. 

When the critical exponent $z=1$, quasiparticles confined inside the expanding critical region have a definite speed of sound $c$. A quasiparticle mode evolves adiabatically under continuous change of its parameters when
\be 
\left| \dot \omega/\omega \right| \ll \omega,
\ee 
where $\omega(t)$ is its instantaneous frequency. For a quasiparticle confined in the gapless region of size $L$, the frequency $\omega\approx c/L$ and the condition becomes
\be 
\dot L \ll c,
\ee 
i.e., the critical region's expansion has to be subsonic.


Setting $v\approx c$ in~\eqref{eq:tau_tot_inhomo}, we obtain an estimate for a lower bound on the total time required for the inhomogeneous adiabatic preparation:
\be 
\tau_{total} \gg \frac{\pi}{\alpha c} + \frac{1}{c} \frac{L}{\sqrt2} \approx \frac{1}{c} \frac{L}{\sqrt2}.
\label{eq:tau_tot_c}
\ee 
For finite $L$, the time offset $\frac{\pi}{\alpha c}$ favors steeper ramps with larger $\alpha$, however, we have to keep in mind that~\eqref{eq:tau_tot_c} requires smooth ramps with $\alpha\ll 1$ such that the lattice discreteness can be ignored. Therefore, within the range of applicability of~\eqref{eq:tau_tot_c}, $\alpha\lessapprox 1$ is the optimal choice, but with increasing $L$, any $\alpha\ll 1$ becomes equally good.
For a smooth ramp and large enough $L$, we have $\tau_{total}\propto L$ that scales with $L$ better than the total time required to make the uniform ramp adiabatic. For $z=1$, the latter is equal to $\pi L^{1+\frac{1}{r\nu}}$ according to~\eqref{eq:tau_adiab} combined with~\eqref{eq:tau_tot_uni}.

\begin{figure}[t!]
\begin{centering}
\includegraphics[width=0.999\columnwidth]{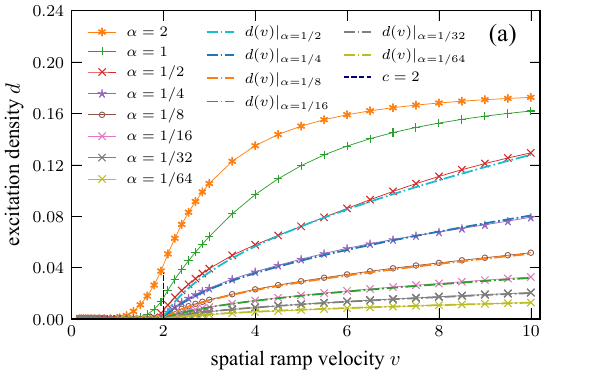}
\includegraphics[width=0.999\columnwidth]{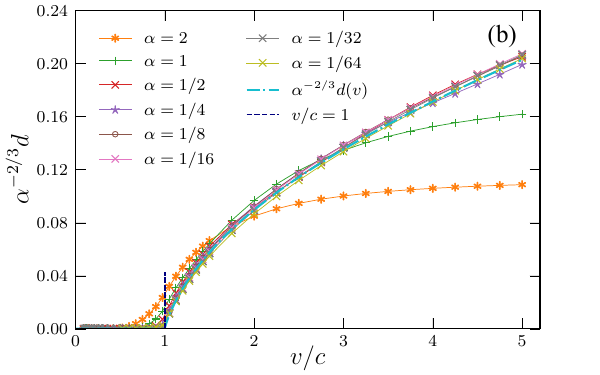}
\includegraphics[width=0.999\columnwidth]{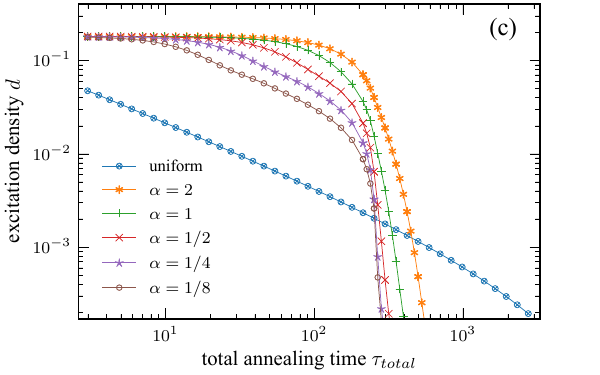}
\par\end{centering}
\caption{ 
{\bf Quantum Ising chain. }
In (a), 
the final density of quasiparticle excitations as a function of the ramp velocity $v$ for several ramp slopes $\alpha$. For smooth ramps with $\alpha\ll 1$, there are no excitations for $v$ below $c=2$. Here, the theoretical prediction for $v>c$ is $d(v)=A_2(\alpha v)^{2/3}(1-c^2/v^2)^{5/6}$, see~\eqref{eq:dv}. $A_2=0.045$ is a coefficient fitted to the data for $\alpha=1/64$.
In (b),
the same data as in (a) but presented as scaled density $\alpha^{-2/3}d$, plotted against scaled velocity $v/c$. For smooth ramps, they collapse to a single plot consistent with scaled $\alpha^{-2/3}d(v)=A_2c^{-2/3}(v/c)^{2/3}(1-c^2/v^2)^{5/6}$.
In (c),
the density as a function of the total preparation time $\tau_{total}$ for the uniform~\eqref{eq:tau_tot_uni} and inhomogeneous~\eqref{eq:tau_tot_inhomo} ramps. The data demonstrate the advantage of the inhomogeneous ramp, here for the chain length $L=1000$.
\label{fig:1D_Ising}
}
\end{figure}

\section{Quantum Ising chain}
\label{sec:QIC}

To begin with, we consider the 1D quantum Ising chain:
\be 
H=-\sum_{s=1}^{L-1} J_{s,s+1}(t)~ \sigma^z_s\sigma^z_{s+1} -\sum_{s=1}^L g_s(t)~ \sigma^x_s .
\label{eq:H_Ising_1D}
\ee 
The transverse field and the interaction are ramped as $g_s(t)=1+\epsilon(u)$ and $J_{s,s+1}=1-\epsilon(u)$, respectively, towards the quantum critical point at $g=1=J$, characterized by the exponents $z=\nu=1$. The speed of quasiparticles at the critical point is $c=2$.

Figure~\ref{fig:1D_Ising}(a) shows the final density of excited quasiparticles in a finite chain with open boundary conditions after an inhomogeneous ramp, as a function of the ramp velocity. For smooth ramps with slope $\alpha\ll1$, we can clearly see suppression of excitations when the ramp is slower than the speed of sound $c=2$. 
The picture is further corroborated in Fig.~\ref{fig:1D_Ising}(b), where the density is scaled by a factor $\alpha^{-2/3}$ and shown as a function of scaled velocity $v/c$. For the smooth ramps, the scaled plots collapse. Setting $s_c=0$, the profile~\eqref{eq:u_inhomo} becomes $u=\alpha(vt-s)$ and, from the point of view of site $s$, it appears locally like a uniform ramp with a ramp time
\be 
\tau=\alpha^{-1}v^{-1}.
\label{eq:tau_local}
\ee 
As we elaborate in Appendix~\ref{app:xx}, for supersonic $v>c$, the ramp indeed proceeds as if it were locally uniform, leaving behind the quasiparticle density scaling like $d\propto \tau_{local}^{-2/3}\propto\alpha^{2/3}$. This effective locality follows from the inability to communicate what happened behind the supersonic front to the area ahead of it. However, the effective local time undergoes a dilatation that decorates the density with an extra quasi-Lorentz factor:
\bea
d(v>c) \approx A_2 \left(\alpha v\right)^{2/3} \left(1-\frac{c^2}{v^2}\right)^{3/4}.
\label{eq:dv_r2}
\eea
Here, $A_2$ is a numerical constant; compare~\eqref{eq:dv} in App.~\ref{app:xx}. A numerical integration similar to the one in the appendix yields $A_2=0.049$, while the best fit in Fig.~\ref{fig:1D_Ising} is $A_2=0.045$. The latter value determines the function $d(v)$ that compares well with the numerical data in Figs.~\ref{fig:1D_Ising}(a,b) for all smooth ramps with the slope $\alpha\ll1$.   

For a uniform ramp, the excitation density is $d\propto\hat\xi^{-1}\propto\tau^{-2/3}$, and the uniform ramp becomes adiabatic when $\hat\xi\gg L$ or, equivalently, $\tau\gg L^{3/2}$. 
Fig.~\ref{fig:1D_Ising}(c) shows the density of excitations as a function of the total ramp time for both the uniform and inhomogeneous ramps. It demonstrates a clear advantage of subsonic smooth inhomogeneous ramps with~$\alpha\ll 1$. 

\section{2D free fermions}
\label{sec:2Dfermions}

A fermionic representation of the integrable 1D Ising chain naturally generalizes to $p$-wave paired spinless fermions on a square lattice:
\be 
H=
\sum_{\langle s, s' \rangle} 
\left(
c_s^\dag c_{s'} - \gamma c_s c_{s'} +{\rm h.c.}
\right) - 
4\lambda \sum_s c_s^\dag c_s.
\ee 
Here, $c_s$ is a fermionic annihilation operator, and for a given $s$, the sum runs over its two NN sites $s'=s+\hat e_x$ and $s'=s+\hat e_y$. The frequency spectrum of fermionic Bogoliubov quasiparticles is
\bea 
\omega(k_x,k_y) 
&=& 2
\left[
\left( \cos k_x + \cos k_y - 2\lambda \right)^2 \right. \nonumber\\
&&
~~~~~~~~~~~~~~
\left.
+\: \gamma^2
\left(
\sin k_x + \sin k_y
\right)^2
\right]^{1/2}.
\eea 
For $\gamma>0$, the model is in a critical gapless $p$-wave superconducting phase when $0<\lambda<1$, and in a gapped superconducting phase when $\lambda>1$. In the following, we set $\gamma=1$ for definiteness. When $\lambda>1$, the gap is 
$
\omega(0,0)=4|\lambda-1|^1.
$ 
Near the phase boundary, it closes with an exponent $z\nu=1$. In the following, we consider ramps ending at $\lambda_f=1$ and parameterized as 
\be 
\lambda(u)=\lambda_f+(\lambda_i-\lambda_f)\epsilon(u),
\ee 
where $\lambda_i=2$ is an initial value. At $\lambda_f=1$, the gap closes at a Fermi point $(k_x,k_y)=(0,0)$.

This Fermi point results from merging, as $\lambda\to1^-$, of two Fermi points that coexist within the gapless phase, $0<\lambda<1$, located at $\pm(k_F,-k_F)/\sqrt2$, where $k_F=\sqrt2\arccos\lambda$. Near each of them, we can parametrize $(k_x,k_y)=(\pm k_F+q+p,\mp k_F-q+p)/\sqrt2$ and expand the dispersion for small $p,q$. Close to both Fermi points, $\omega^2\approx c^2_p(\lambda)p^2 + c^2_q(\lambda) q^2$, where $c_p$ and $c_q$ are the quasiparticle velocities in orthogonal directions. $c_p(\lambda)$ monotonically increases with $\lambda$ from $c_p(0)=0$, while $c_q(\lambda)$ decreases to $c_q(1)=0$ at the phase boundary. At $\lambda=1$, the two Fermi points merge at $(k_x,k_y)=(0,0)$. Near this double Fermi point, there is a finite speed 
\be 
c_p=2\sqrt2
\label{eq:cp}
\ee
in one direction and a quadratic dispersion $\omega\propto q^2$ along the orthogonal direction. The dynamical exponents in these directions are $z=1$ and $z=2$, respectively. The quadratic direction is the one along which the two Fermi points merge as $\lambda\to1^-$.

\begin{figure}[t!]
\begin{centering}
\includegraphics[width=0.99\columnwidth]{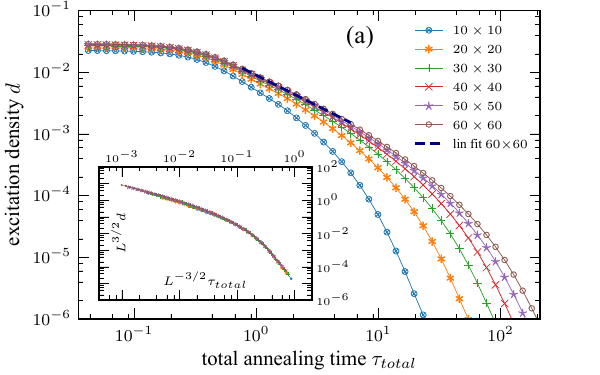}
\includegraphics[width=0.99\columnwidth]{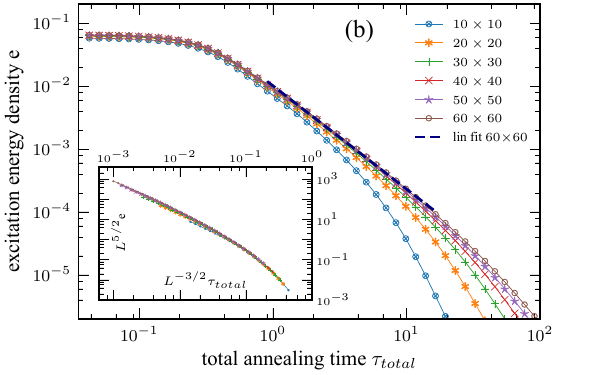}
\par\end{centering}
\caption{ 
{\bf 2D fermions - uniform ramp. }
In (a), the density of excited quasiparticles $d$ as a function of the total ramp time $\tau_{total}$ for different system sizes $L\times L$. $d\propto\tau^{-1}$ in~\eqref{eq:d_uni_2DF} is consistent with the data for large enough system sizes. The fit for $L=60$ in the linear region gives exponent $-0.99(3)$ with uncertainties corresponding to the $95\%$ confidence interval.
In (b), the density of excitation energy, ${\rm e}$, as a function of the ramp time. ${\rm e}\propto \tau^{-5/3}$ in~\eqref{eq:e_uni_2DF} is consistent with the data for large enough $L$, where the fit for $L=60$ gives exponent $-1.64(4)$.
The insets in (a) and (b) show, respectively, scaled $L^{3/2}d$ and $L^{5/2}{\rm e}$ as functions of scaled time $L^{-3/2}\tau_{total}$. The collapse of the scaled plots confirms that $\tau \gg L^{3/2}$ marks the crossover to the adiabatic regime.
\label{fig:2D_fermions_uniform}
}
\end{figure}

\subsection{Uniform ramp}
\label{sec:2Dfermions-uniform}

For $\lambda_f=1$, the uniform ramp approaches the critical point in a parabolic way with $r=2$, and the gap closes with $z\nu=1$. It fails to be adiabatic at $\hat t \propto \tau^{rz\nu/(1+rz\nu)} = \tau^{2/3}$. In each of the two orthogonal directions with $z=1,2$, quasiparticles become excited up to $\hat k_z=(\hat t/\tau)^{r\nu}=\tau^{-2/(3z)}$ and, therefore, their density should scale with the ramp time as
\be 
d \propto \hat k_1 \hat k_2 \propto \tau^{-1} .
\label{eq:d_uni_2DF}
\ee 
This power law is consistent with the numerical results in Fig.~\ref{fig:2D_fermions_uniform} for sufficiently large system size $L$. 

For a crosscheck, we also consider the excitation energy per site. At $\lambda_f=1$, the dispersion near the Fermi point is
\bea 
\omega^2 
& \approx & 
8q^2 + p^4 - \frac{q^4}{3} -2 q^2p^2  \label{eq:disp_aniso} \\
& \approx &
\tau^{-4/3}
\left[
8 a_1^2 \left(\frac{q}{\hat k_1}\right)^2 + a_2^4 \left(\frac{p}{\hat k_2}\right)^4
\right]. \label{eq:disp_aniso_largetau}
\eea 
In the second line, we kept only the leading term for large $\tau$. Here, the non-universal constants $a_{1,2}$ are coefficients in the asymptotes $\hat k_1=a_1\tau^{-2/3}$ and $\hat k_2=a_2\tau^{-1/3}$. Given that $|p|$ and $|q|$ are excited in the range from $0$ up to $\hat k_1$ and $\hat k_2$, respectively, Eq.~\eqref{eq:disp_aniso_largetau} implies average energy of excited quasiparticles $\propto \tau^{-2/3}$ and the excitation energy density scaling as
\be 
{\rm e} \propto \tau^{-2/3} ~ d \propto \tau^{-5/3},
\label{eq:e_uni_2DF}
\ee
in agreement with Fig.~\ref{fig:2D_fermions_uniform}(b).

\begin{figure}[t!]
\begin{centering}
\includegraphics[width=0.999\columnwidth]{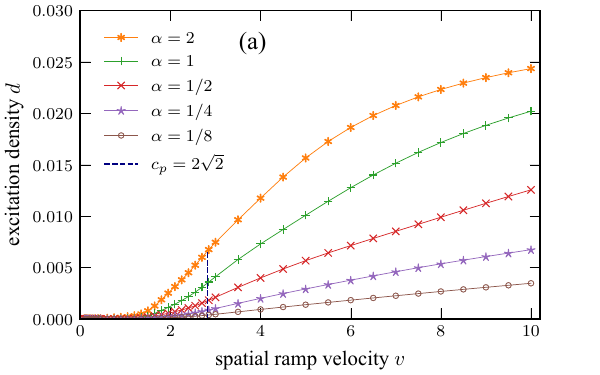}
\includegraphics[width=0.999\columnwidth]{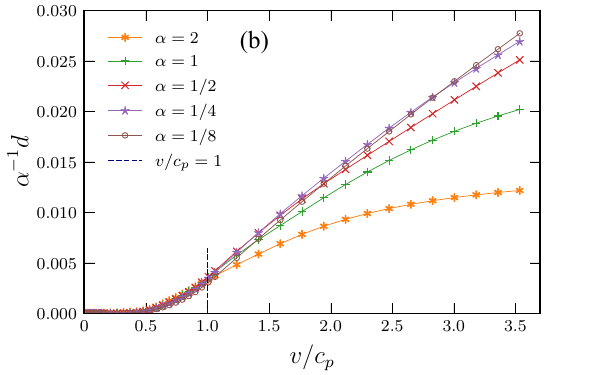}
\par\end{centering}
\caption{ 
{\bf 2D fermions - effective speed limit. }
In (a), the final density of quasiparticle excitations as a function of the ramp velocity $v$.
The ramp ends at $\lambda_f=1$ where the speed is highly anisotropic: $c_p=2\sqrt{2}$ but $c_q=0$. Here $c_p$ is marked by a vertical dashed line.
For the smooth ramps with $\alpha\ll1$ and the fastest velocity, the density satisfies $d\propto\alpha^{1}$ in accordance with the power laws for uniform ramps in Fig.~\ref{fig:2D_fermions_uniform}. A smooth supersonic ramp is effectively uniform.
In (b), the same data as in (a), but presented as scaled density $\alpha^{-1}d$ as a function of $v/c_p$. 
For decreasing $\alpha$, they tend to collapse to a single plot. 
Here, the lattice size $60\times 60$ is much larger than the ramps' widths $1/\alpha$. 
\label{fig:2D_fermions_speed}
}
\end{figure}

For large enough $L$, Figs.~\ref{fig:2D_fermions_uniform}(a) and ~\ref{fig:2D_fermions_uniform}(b) demonstrate the crossover to the adiabatic regime. In a finite system of linear size $L$, the minimal quasimomenta $p,q\approx 1/L$ and, for large enough $L$, it is the nonzero speed $c_p=2\sqrt2$ in the direction $z=1$ that determines the minimal gap $\propto L^{-1}$ in~\eqref{eq:disp_aniso}. Accordingly, the KZ scale of length that is relevant for the minimal gap is $\hat\xi_1\propto \hat k_1^{-1} \propto \tau^{2/3}$, and the uniform ramp is adiabatic when $\hat\xi_1\gg L$ or, equivalently, $\tau \gg L^{3/2}$. This prediction agrees with the insets in Figs.~\ref{fig:2D_fermions_uniform}(a) and \ref{fig:2D_fermions_uniform}(b) that show, respectively, scaled $L^{3/2}d$ and $L^{5/2}{\rm e}$ as functions of scaled time $L^{-3/2}\tau_{total}$. The collapse of the scaled plots confirms that $\tau \propto L^{3/2}$ marks the crossover to the adiabatic regime.

\begin{figure}[t!]
\begin{centering}
\includegraphics[width=0.99\columnwidth]{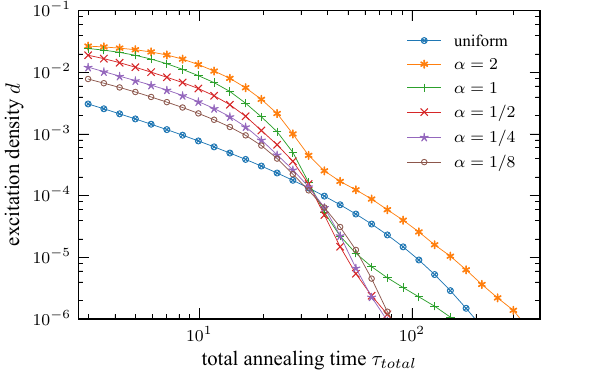}
\par\end{centering}
\caption{ 
{\bf 2D fermions - total time. }
The final density of quasiparticle excitations as a function of total preparation time $\tau_{total}$ for uniform~\eqref{eq:tau_tot_uni} and inhomogeneous~\eqref{eq:tau_tot_inhomo} ramps. 
The data demonstrate the advantage of the inhomogeneous ramp with a suitable slope $\alpha$.
Here, the lattice size $60\times 60$ is much larger than the ramp's width $1/\alpha$. 
\label{fig:2D_fermions_total}
}
\end{figure}

\subsection{Inhomogeneous ramp}
\label{sec:2Dfermions-inhomo}

The dispersion near the anisotropic double Fermi point~\eqref{eq:disp_aniso} has a finite speed $c_p=2\sqrt2$ along $p$ and a quadratic dispersion along $q$. In an expanding critical region of linear size $a$, the minimal quasimomenta are $p,q\approx 1/a$ and they determine the minimal frequency
\be 
\omega_m = c_p a^{-1} + {\cal O}\left( a^{-3} \right).
\ee 
For large enough $a$, the adiabatic condition, $|\dot\omega_m/\omega_m| \ll \omega_m$, becomes
\be 
\frac{da}{dt} \ll c_p.
\ee 
Despite the anisotropy, $c_p$ marks a crossover to a regime where the evolution is adiabatic (see Fig.~\ref{fig:2D_fermions_speed}).
As far as causality is concerned, the two velocities $c_p=2\sqrt2$ and $c_q=0$ in the orthogonal directions imply that in almost any direction (except direction $q$ that is a set of measure zero), there is a finite velocity $\propto c_p$ and, therefore, $c_p$ serves as a velocity scale that marks a crossover between supersonic and subsonic/adiabatic.   

As a self-consistency check, we notice that for the smooth ramps $\alpha=1/2,1/4,1/8$ and the fastest velocity in Fig.~\ref{fig:2D_fermions_speed}(a), the density $d$ is linear in $\alpha$. When we fix the position $d(s,s_c)$ in the inhomogeneous~\eqref{eq:u_inhomo} and compare it with the uniform~\eqref{eq:u_uni}, we find that locally the ramp proceeds with a ramp time $\tau=\alpha^{-1}v^{-1}$. This formula translates $d\propto\tau^{-1}$, valid for a uniform ramp, to $d\propto\alpha$ for a supersonic inhomogeneous ramp; compare with the scaled plots in Fig.~\ref{fig:2D_fermions_speed}(b). For a supersonic ramp, excited quasiparticles cannot propagate across the ramp and, for this lack of communication, the ramp proceeds as if it were locally uniform.

The subsonic adiabaticity results in the inhomogeneous advantage demonstrated in Fig.~\ref{fig:2D_fermions_total}, where the density of excitations is shown as a function of the total ramp time for several values of the slope $\alpha$. The smooth ramps with $\alpha\lessapprox 1$ reduce the density below that of the uniform ramp for a sufficiently long total time. In other words, when a low density of excitations is desired then it can be achieved by an inhomogeneous ramp faster than by a uniform one.

\begin{figure}[t!]
\begin{centering}
\includegraphics[width=0.99\columnwidth]{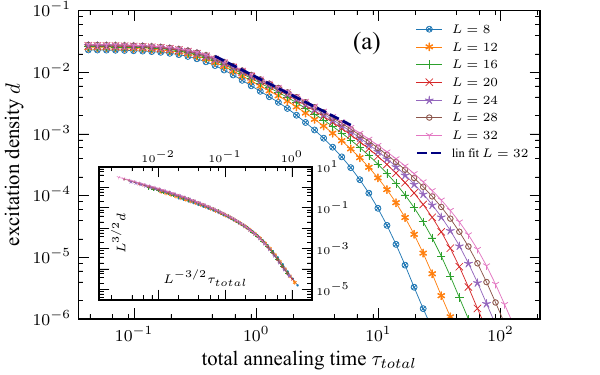}
\includegraphics[width=0.99\columnwidth]{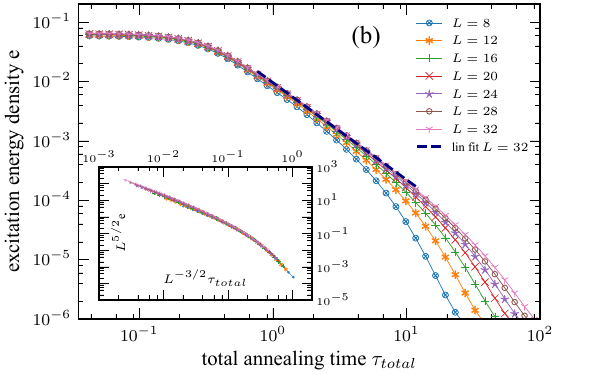}
\par\end{centering}
\caption{ 
{\bf Kitaev model - uniform ramps. }
In (a), the density of excited quasiparticles for a ramp ending at $\lambda_f=2$ for different system sizes $L$, where the lattice is a hexagon with $L$ hexagonal plaquettes at its edge, see the inset in Fig.~\ref{fig:Kitaev_total}. 
The data are consistent with $d\propto \tau^{-1}$, where the fit for $L=32$ in the linear region gives exponent $-1.00(3)$ with uncertainty at $95\%$ confidence interval.
In (b), the excitation energy per site. The data are consistent with ${\rm e}\propto \tau^{-5/3}$, where the fit for $L=32$ gives exponent $-1.64(4)$.
\label{fig:Kitaev_uniform}
}
\end{figure}

\begin{figure}[t!]
\begin{centering}
\includegraphics[width=0.99\columnwidth]{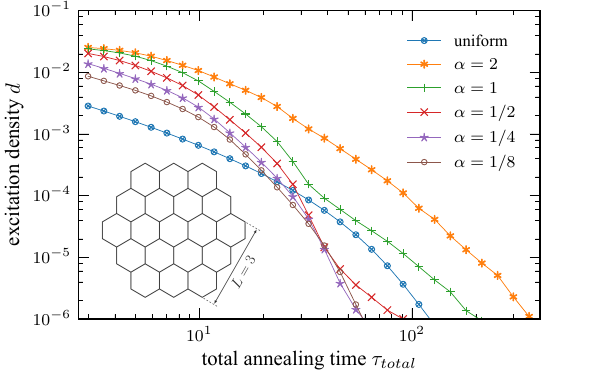}
\par\end{centering}
\caption{ 
{\bf Kitaev model - total time. }
The final density of quasiparticle excitations as a function of total preparation time $\tau_{total}$ for a uniform~\eqref{eq:tau_tot_uni} and an inhomogeneous~\eqref{eq:tau_tot_inhomo} ramp. 
The data demonstrate the advantage of the inhomogeneous ramp for slope $\alpha<1$.
Here, the lattice is a hexagon with $L=32$ hexagonal plaquettes at its edge. 
The inset shows a hexagon with $L=3$ as an example.
\label{fig:Kitaev_total}
}
\end{figure}

\begin{figure}[t!]
\begin{centering}
\includegraphics[width=0.999\columnwidth]{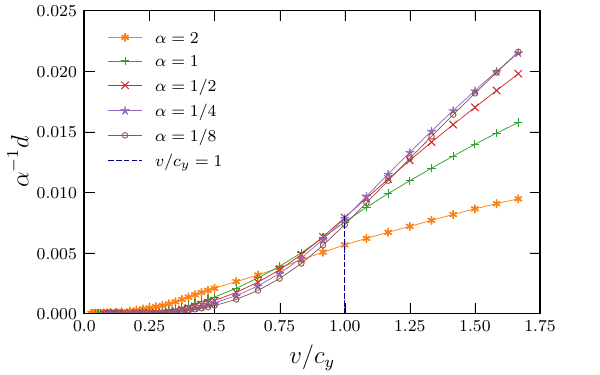}
\par\end{centering}
\caption{ 
{\bf Kitaev model - effective speed limit. }
The scaled density of quasiparticle excitation as a function of the scaled velocity of the inhomogeneous ramp as in Fig.~\ref{fig:2D_fermions_speed}. The plots tend to collapse for smooth ramps, $\alpha\ll1$.
Here, the lattice is a hexagon with $L=32$.
\label{fig:Kiatev_speed}
}
\end{figure}

\section{ Kitaev model }
\label{sec:kitaev}

The Kitaev Hamiltonian~\cite{kitaev2003} in a vortex-free sector becomes~\cite{KZ_Kitaev}
\be 
H_K=i\sum_{\vec n} 
b_{\vec n}
\left( 
J^x_{\vec n} a_{\vec n - \vec M_1} + J^y_{\vec n} a_{\vec n +\vec M_2} + J^z_{\vec n} a_{\vec n} \right).
\label{eq:Hab}
\ee 
Here, $\vec n$ numbers the (vertical) $J^z$ bonds of the hexagonal lattice. $a_{\vec n}$ and $b_{\vec n}$ are Majorana fermions sitting at the top and bottom ends of the $\vec n$-th $J^z$ bond, respectively. They can be expressed by
\bea 
b_{\vec n}=c_{\vec n}+c_{\vec n}^\dag,~~~~ i a_{\vec n}=c_{\vec n}-c_{\vec n}^\dag,
\eea 
where the fermionic annihilation operators $c_{\vec n}$ live on a square lattice. The Hamiltonian~\eqref{eq:Hab} is quadratic in $c_{\vec n}$ and $c_{\vec n}^\dag$, and the dispersion of quasiparticle excitations in a uniform infinite system reads
\be 
\omega(k_x,k_y)=
2
\left| 
J_z + J_x e^{ i\vec k \vec M_1 } + J_y e^{ -i \vec k \vec M_2}
\right|,
\label{eq:disp_K}
\ee
where $\vec M_1 = (\sqrt3/2,3/2)$ and $\vec M_2 = (\sqrt3/2,-3/2)$. 

Here, we fix $J^x=J^y=1$ for definiteness, making the gapless critical phase confined to $0\leq J^z\leq 2$, and consider a ramp of $J^z_{\vec n}$ from $J_i=4$ in the gapfull phase to $J_f=2$ at the boundary of the gapless one:
\be 
J^z_{\vec n}=J_f+(J_i-J_f)\epsilon(u).
\ee 
The relevant properties of the spectrum~\eqref{eq:disp_K} are the same as those of the free fermions. At $J^z=2$, there is a double Fermi point at $\left(0,2\pi/3\right)$ with dynamical exponents $z=1,2$ depending on the direction. It has a quadratic dispersion along $k_x$ and a linear one along $k_y$ with velocity $c_y=6$. 
Consequently, we obtain the same power laws in a uniform transition and a similar inhomogeneous advantage as for the paired fermions, see Figs.~\ref{fig:Kitaev_uniform} and~\ref{fig:Kitaev_total}, with the advantage stemming from an effective speed limit set by $c_y$ despite vanishing $c_x$, see Fig.~\ref{fig:Kiatev_speed}.

\begin{figure}[t!]
\begin{centering}
\includegraphics[width=0.99\columnwidth]{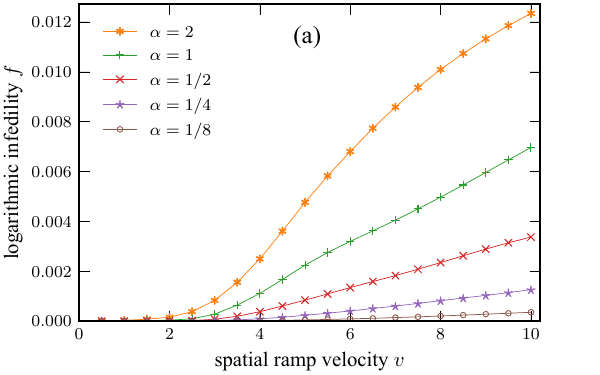}
\includegraphics[width=0.99\columnwidth]{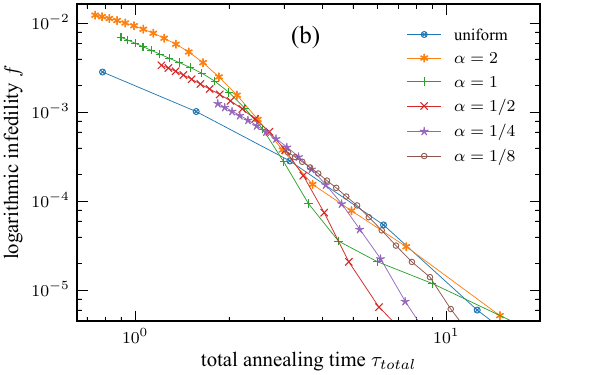}
\par\end{centering}
\caption{ 
{\bf 2D quantum Ising model. }
In (a), the logarithmic infidelity~\eqref{eq:f} as a function of the velocity $v$ for different slopes $\alpha$.
In (b), the log-infidelity as a function of the total time.
Here lattice size is $10\times10$, maximum bond dimension of a matrix product state $\chi=128$, and time step $dt=0.005$. 
\label{fig:Ising_2D}
}
\end{figure}

\section{2D quantum Ising model}
\label{sec:2D_QIM}

The last example considered is the (nonintegrable) quantum Ising model on a square lattice:
\be 
H = - \sum_{\langle s,s' \rangle} J_{ss'} \sigma^z_s \sigma^z_{s'} - \sum_s g_s \sigma^x_s .
\label{eq:H_Ising_2D}
\ee 
The ramp is implemented as
$
g_s(t)     = g_c [ 1 + \epsilon(u) ]
$
and
$
J_{ss'}(t) = J_c [ 1 - \epsilon(u) ]
$, 
with $u$ being either uniform~\eqref{eq:u_uni} or inhomogeneous~\eqref{eq:u_inhomo}. We fix the energy scale by setting $J_c = 1$. The para-ferromagnetic quantum phase transition is located at $g_c/J_c= 3.04438$~\cite{Deng_QIshc_02}. 

A quantity of interest is the logarithmic infidelity between the final state and the final adiabatic ground state:
\be 
f=-\frac{1}{N}\ln \left| \langle \psi_f | GS_f \rangle \right|^2,
\label{eq:f}
\ee
where $N$ is the number of sites. In an integrable fermionic model, we would have
\be 
f=-\frac{1}{N} \sum_j \ln(1-p_j) \approx \frac{1}{N} \sum_j p_j = d.
\ee 
Here, the sum runs over all fermionic quasiparticle modes of the final Hamiltonian, $p_j$ is the excitation probability of the $j$-th mode, and the approximation is accurate when $p_j\ll1$ for all modes. Therefore $f$ can be thought of as a substitute for the density of quasiparticle excitations $d$ in the case of a nonintegrable system.

Our simulation results are shown in Fig.~\ref{fig:Ising_2D}. The exponential growth of the matrix product state's bond dimension with system size limits the simulations to a $10\times10$ lattice with open boundary conditions.
Figure~\ref{fig:Ising_2D}(a) shows the infidelity as a function of the ramp velocity. It is clearly suppressed at low velocities but the speed of sound is obscured by finite size effects.
Nevertheless, as shown in Fig.~\ref{fig:Ising_2D}(b), there is a clear advantage of a suitable inhomogeneous ramp over the homogeneous one in terms of the infidelity as a function of the total preparation time. Due to the limited size, $\alpha=1,1/2$ are the optimal slopes as they not only minimize the excitation but also their small width limits the time offset.   

\section{Conclusion}
\label{sec:conclusion}

For a system with a definite speed of sound at the critical point, $c$, a subsonic inhomogeneous ramp starting from the center of the lattice requires shorter adiabatic preparation time than a ramp uniform in space.

This conclusion generalizes to situations where the energy gap is inversely proportional to the linear size of the expanding critical region. The numerical prefactor of the proportionality defines a velocity scale that marks a crossover to the adiabatic regime. 

The data used for the figures in this article are openly available from the RODBUK repository at Ref.~\onlinecite{UJ/UNMTRB_2024}.

\acknowledgements
This research was 
funded by the National Science Centre (NCN), Poland, under project 2021/03/Y/ST2/00184 within the QuantERA II Programme that has received funding from the European Union’s Horizon 2020 research and innovation programme under Grant Agreement No 101017733 (FB,IS)
and NCN under projects 2019/35/B/ST3/01028 (JD) and 2020/38/E/ST3/00150 (MMR).
%
\bibliography{KZref.bib}
\appendix

\section{A supersonic ramp in the 1D quantum Ising model}
\label{app:xx}

For clarity and to connect with Refs.~\onlinecite{d2005,inhomo_quantum-a}, in this appendix, we set $J=1$ in the Hamiltonian~\eqref{eq:H_Ising_1D} and ramp only the transverse field $g_s$. Furthermore, we assume 
that an inhomogeneous ramp does not start in the center but is moving from the left to right with velocity $v$. Behind the ramp, at site numbers $s<vt$, the transverse field is the critical $g_s=1$ and for $s>vt$, it increases monotonically to $g_s=1+\epsilon_i$ far ahead of the ramp. The critical $g_s=1$ is approached as 
\be 
\epsilon_s =
g_s-1 =
\epsilon\left[\alpha(s-vt)\right] \approx 
\epsilon_0 \cdot \left[\alpha (s-vt)\right]^r
\label{eq:eps_s}
\ee 
for sufficiently small positive $\alpha(s-vt)$. For $v$ greater than the speed of sound at the critical point, $c=2$, we will map the inhomogenous ramp to an equivalent uniform one.

After the Jordan-Wigner transformation to spinless fermionic operators $c_n$, 
\bea
\sigma^x_s &=& 1-2 c^\dagger_s c_s,\\
\sigma^z_s &=& -\left( c_s + c_s^\dagger \right)\prod_{s'<s}(1-2 c^\dagger_{s'} c_{s'} ),
\eea 
the Hamiltonian becomes 
\bea
H ~=~ 2 \sum_{s=1}^L g_s ~c_s^\dagger  c_s-
\sum_{s=1}^{L-1}
\left( 
c_s^\dag c_{s+1} + c_{s+1} c_s + {\rm h.c.}
\right)
\label{Hpm}
\eea
It is diagonalized to $H=E_0+\sum_m\omega_m\gamma_m^\dag\gamma_m$ by a Bogoliubov transformation $c_s=\sum_{m=1}^{L}(u_{s,m}\gamma_m + v^*_{s,m} \gamma_m^\dagger)$ with $m$ numerating $L$ eigenmodes of stationary Bogoliubov-de Gennes (BdG) equations,
\bea
\omega_m u_{s,m}^\pm = 2 g_s u_{s,m}^\mp - 2 u_{s\mp 1,m}^\mp,
\label{eq:BdG}
\eea
with $\omega_m\geq0$. Here, $u_{s,m}^\pm\equiv u_{s,m} \pm v_{s,m}$. A time-dependent version of the BdG equations is
\bea
i\partial_t 
\left(
\begin{array}{c}
u^+_s \\
u^-_s
\end{array}
\right)
&=&
2 \epsilon\left[\alpha(s-vt)\right] \sigma^x 
\left(
\begin{array}{c}
u^+_s \\
u^-_s
\end{array}
\right) + \nonumber\\
& &
2 \sigma^x 
\left(
\begin{array}{c}
u^+_s - u^+_{s+1} \\
u^-_s - u^-_{s-1}
\end{array}
\right)~.
\label{eq:tBdG}
\eea
Here, we suppressed the mode number $m$. For a uniform ramp, we would have $\epsilon=g_s-1\propto\left(|t|/\tau\right)^r$ for sufficiently small negative $t/\tau$ in place of the inhomogeneous~\eqref{eq:eps_s}. Locally, i.e., for a fixed position $s$, the inhomogenous~\eqref{eq:eps_s} approaches zero on a time scale
\be 
\tau\propto(\alpha v)^{-1}.
\label{eq:tau_local}
\ee 
For a supersonic ramp, this simple observation can be refined by mapping the inhomogeneous (\ref{eq:tBdG}) to a uniform transition with an effective ramp time $\tilde\tau>\tau$. 

To this end, we make a transformation:
\bea
s &=& \tilde s, 
\label{eq:primes} \\
t &=& \frac{\tilde s}{v} + \left(1-\frac{c^2}{v^2}\right) \tilde t.
\label{eq:primet}
\eea
It introduces a local time variable $\tilde t$ measured with respect to time when $\epsilon_s$ becomes $0$ for a given $s$. Defining $\gamma\equiv\sqrt{1-c^2/v^2}$ and 
\bea 
u^{\pm}_s(t) =
u^{\pm}_{\tilde s}\left(\frac{\tilde s}{v}+\gamma^2\tilde t\right) \equiv
\tilde u_{\tilde s}^\pm \left(\tilde t\right),
\eea
we obtain 
\bea 
&&
u^+_{s+1}(t) =
u^+_{\tilde s +1}\left(\frac{\tilde s}{v}+\gamma^2\tilde t\right) =
\tilde u^+_{\tilde s +1}\left(\tilde t-\frac{1}{v\gamma^2}\right), \\
&&
u^-_{s-1}(t) =
u^-_{\tilde s -1}\left(\frac{\tilde s}{v}+\gamma^2\tilde t\right) =
\tilde u^-_{\tilde s -1}\left(\tilde t+\frac{1}{v\gamma^2}\right).
\eea 
and~\eqref{eq:tBdG} becomes
\bea
i \partial_{\tilde t} 
\left[
\begin{array}{c}
\tilde u^+_{\tilde s}(\tilde t) \\
\tilde u^-_{\tilde s}(\tilde t)
\end{array}
\right]
&=&
2 \gamma^{2} \epsilon \left( \alpha\gamma^2v\tilde t\right ) \sigma^x 
\left[
\begin{array}{c}
\tilde u^+_{\tilde s}(\tilde t) \\
\tilde u^-_{\tilde s}(\tilde t)
\end{array}
\right] + \nonumber\\
& &
2 \gamma^{2} \sigma^x 
\left[
\begin{array}{c}
\tilde  u^+_{\tilde s}(\tilde t) - \tilde u^+_{\tilde s+1} \left(\tilde t-\frac{1}{v\gamma^2}\right) \\
\tilde  u^-_{\tilde s}(\tilde t) - \tilde u^-_{\tilde s-1} \left(\tilde t+\frac{1}{v\gamma^2}\right)
\end{array}
\right]~.
\label{eq:tildetBdG}
\eea
Anticipating that relevant modes are smooth in $\tilde s$, we expand $\tilde u^{\pm}_{\tilde s\pm1}$ to first order in $\partial_{\tilde s}$ and then also to first order in $\partial_{\tilde t}$:
\bea
&&
i \left( 1 + \frac{c}{v} \sigma^y + \frac{c}{v} \sigma^x i\partial_{\tilde s} \right)
\partial_{\tilde t}
\left(
\begin{array}{c}
\tilde u^+ \\
\tilde u^-
\end{array}
\right) = \nonumber\\
&&
2 \gamma^{2} \epsilon \left( \alpha\gamma^2v\tilde t\right ) \sigma^x 
\left(
\begin{array}{c}
\tilde u^+ \\
\tilde u^-
\end{array}
\right) + 
2 \gamma^{2} \sigma^y i\partial_{\tilde s}
\left(
\begin{array}{c}
\tilde  u^+ \\
\tilde  u^-
\end{array}
\right)~.
\label{eq:tildetBdG}
\eea
Here, $\tilde u^{\pm}=\tilde u^\pm_{\tilde s}(\tilde t)$ for short.
A substitution
\bea
\left(
\begin{array}{c}
\tilde u^+\\
\tilde u^-
\end{array}
\right) =
S_v
\left(
\begin{array}{c}
a_{\tilde k}\\
b_{\tilde k}
\end{array}
\right)
e^{i\tilde k\tilde s}
~,
\label{eq:Sv}
\eea
where
\be 
S_v=
\left(
\begin{array}{cc}
\sqrt{1-\frac{c^2}{v^2}} & i\frac{c}{v}  \\
0                        & 1
\end{array}
\right)
\ee
brings~\eqref{eq:tildetBdG} to  
\bea
i\partial_{\tilde t}
\left(
\begin{array}{c}
a_{\tilde k} \\
b_{\tilde k}
\end{array}
\right)
&=&
2\gamma 
\epsilon\left(\alpha\gamma^2 v\tilde t\right) 
\sigma^x
\left(
\begin{array}{c}
a_{\tilde k} \\
b_{\tilde k}
\end{array}
\right) -
2\tilde k \sigma^v 
\left(
\begin{array}{c}
a_{\tilde k} \\
b_{\tilde k}
\end{array}
\right) + \nonumber\\
&&
2\tilde k \frac{c}{v} \left[ 1 + \epsilon\left(\alpha\gamma^2 v\tilde t\right) \right]
\left(
\begin{array}{c}
a_{\tilde k} \\
b_{\tilde k}
\end{array}
\right)~,
\label{eq:ab}
\eea
where we systematically neglected terms $\propto\tilde k^2$ and higher. Here, $\sigma^v = \sigma^y\sqrt{1-\frac{c^2}{v^2}} + \frac{c}{v}~\sigma^z$ is a rotated Pauli matrix, and the second line generates a mere phase factor that can be removed by a redefinition: 
\be
a_{\tilde k}=\tilde{a}_{\tilde k}e^{-i\tilde k\varphi},~~~~
b_{\tilde k}=\tilde{b}_{\tilde k}e^{-i\tilde k\varphi},
\ee
where
\be 
\frac{d\varphi_{\tilde k}}{dt}=
2\frac{c}{v} \left[ 1 + \epsilon\left(\alpha\gamma^2 v\tilde t\right) \right].
\ee
With the asymptote in~\eqref{eq:eps_s} and a dilated ramp time 
\be 
\tilde\tau ~=~\tau~\left(1-\frac{c^2}{v^2}\right)^{-\frac{2r+1}{2r}}
\ee 
\eqref{eq:ab} can be brought to a nonlinear Landau-Zener form~\cite{LZ_nlin}
\be
i \frac{d}{dz}
\left(
\begin{array}{c}
\tilde{a}_{\tilde k}\\
\tilde{b}_{\tilde k}
\end{array}
\right)~=~
\frac12
\left[
~|\delta z|^r ~\sigma^x~-~\sigma^v~{\rm sign}(\tilde k)
\right]
\left(
\begin{array}{c}
\tilde{a}_{\tilde k}\\
\tilde{b}_{\tilde k}
\end{array}
\right)~.
\label{eq:LZ_nlin}
\ee
Here, 
$
z=4|\tilde k|\tilde t
$ 
is a new timelike variable and 
$
\delta=4^{-1}\epsilon_0^{1/r}|\tilde k|^{-(r+1)/r}\tilde\tau^{-1}
$ 
is a transition rate. As $\delta$ is the only parameter, the ramp ends at $z=0$, precisely at the anti-crossing, with a transition probability $p_r(\delta)$. The probability does not depend on the sign of $\tilde k$. It is non-negligible when the transition rate $\delta\gg 1$ or, equivalently, for $|\tilde k|$ up to approximately $\tilde\tau^{-r/(1+r)}$. We note that for $v\to c^+$, the rate $\delta$ shrinks to zero, suggesting adiabaticity for subsonic ramps. In order to obtain the number of excited quasiparticles for $v>c$, the probability $p_r(\delta)$ has to be integrated over quasiparticle modes with a proper measure that can be determined by initial conditions. 

The initial ground state ahead of the supersonic ramp is not affected by the approaching ramp. It is a vacuum for Bogoliubov quasiparticles in the quasimomentum representation with Bogoliubov coefficients: 
\be 
\left(
\begin{array}{c}
u^+\\
u^-
\end{array}
\right)
\propto 
e^{iks}
e^{-it\omega_k}.
\label{eq:u_i}
\ee
Here, $\omega_k$ is a quasiparticle spectrum for the initial constant value of $\epsilon$. If there were any quasiparticles excited above this ground state, their density would be given by an integral,
\be 
\int_{-\pi}^\pi \frac{dk}{2\pi} p_k,
\ee 
averaging excitation probability $p_k$ over the first Brillouin zone with a uniform measure $\frac{dk}{2\pi}$. Every mode~\eqref{eq:u_i} is modified behind the passing ramp, but as it remains orthogonal to the other modes, the measure is preserved. 
In order to express it by $\tilde k$, we note that an inverse of the transformation in~\eqref{eq:primes}, \eqref{eq:primet} and~\eqref{eq:Sv} maps the initial state~\eqref{eq:u_i} to
\be 
\left(
\begin{array}{c}
\tilde u^+\\
\tilde u^-
\end{array}
\right) 
\propto 
e^{i\tilde k\tilde s}
e^{-i \omega_k \tilde t \left(1-c^2/v^2\right) },
\ee
where
$
\tilde k=k-\frac{\omega_k}{v}
$.
For a sufficiently large initial $\epsilon_i$, the quasiparticle dispersion $\omega_k$ is nearly flat, and $\tilde k$ is just shifted by a constant with respect to $k$. Therefore the density of excited quasiparticles can be approximated by an integral over $\tilde k$:
\bea
d(v>c)
& \approx &
\int \frac{d\tilde k}{2\pi}~p_r\left(\delta\right)~ = 
A_r \tilde\tau^{-r/(1+r)} \nonumber\\
& = &
A_r \tau^{-\frac{r}{1+r}} \left(1-\frac{c^2}{v^2}\right)^{\frac{2r+1}{2r+2}} .
\label{eq:dv}
\eea
Here, $A_r$ is a nonuniversal constant. Numerical integration yields $A_2=0.039$, while $A_2=0.029$ is the best fit to data obtained by simulation of a supersonic inhomogeneous ramp of the transverse field $g_s$. We note that the values of $A_2$ provided here are for linear ramp
of the transverse field considered in the Appendix, and the values of $A_2$ appearing in the main text are for a smooth ramp of both transverse field and couplings considered there. 
Equation~\eqref{eq:dv} is used in Fig.~\ref{fig:1D_Ising} as a theoretical function with $A_2$ being its only fitting parameter. Finally, we notice that for $v\gg c$, the density
\be
d(v\gg c) \approx A_r \tau^{-\frac{r}{1+r}}
\ee 
with a local ramp time $\tau$ in~\eqref{eq:tau_local}, is the same as the density after a uniform ramp with the same global $\tau$. 

\end{document}